\documentclass[12pt]{article}
\usepackage{epsfig}
\usepackage{amsbsy}
\usepackage{amsfonts}
\usepackage{amssymb}
\usepackage{amsmath}
\usepackage{graphicx}

\def\id{\protect{{1 \kern-.28em {\rm l}}}}

\def\lin{{-\!\!\!-\!\!\!\!-\!\!\!\!-\!\!\!-\!\!\!-}}

\makeatletter
\renewcommand\section{\@startsection {section}{1}{\z@}%
                                   {-3.5ex \@plus -1ex \@minus -.2ex}%
                                   {2.3ex \@plus.2ex}%
                                   {\normalfont\large\bfseries}}
\renewcommand\subsection{\@startsection{subsection}{2}{\z@}%
                                   {-3.25ex\@plus -1ex \@minus -.2ex}%
                                   {1.5ex \@plus .2ex}%
                                   {\normalfont\normalsize\bfseries}}
\makeatother

\begin{document}

${}$

\vspace{.5cm} 

\begin{center}

%
%
%
%

{\Large  Magnon Bound-state Scattering in Gauge and String Theory  }

\bigskip
\bigskip

 { 
Radu~Roiban 
}
 
\bigskip
\medskip

{
\em Department of Physics, The Pennsylvania  State University,\\
University Park, PA 16802 , USA \\
{\rm radu@phys.psu.edu}
}

\end{center}

\bigskip
\bigskip

\begin{abstract}

It has been shown that, in the infinite length limit, 
the magnons of the gauge theory spin chain can form bound states carrying one finite and 
one strictly infinite R-charge. These bound states have been argued to be associated to simple 
poles of the multi-particle scattering matrix and to world sheet solitons carrying the same 
charges. Classically, they can be mapped  to the solitons of the complex sine-Gordon theory.

Under relatively general assumptions we derive the condition that
simple poles of the two-particle  scattering matrix correspond to
physical bound states and construct higher bound states  ``one magnon at a time''. 
We construct the scattering matrix of the bound states of the BDS and
the AFS S-matrices.  The bound state S-matrix exhibits simple and
double poles
and thus its analytic structure 
is much richer than that of the elementary magnon S-matrix. We also discuss
the bound states  appearing in larger sectors and their S-matrices.
The large 't~Hooft coupling limit of the scattering phase of the 
bound states in the $SU(2)$ sector
is found to agree with the semiclassical scattering of world sheet
solitons. Intriguingly, the contribution of the dressing phase  has an
independent world sheet interpretation
as the soliton-antisoliton scattering phase shift. The small momentum
limit provides independent tests of these identifications.

\end{abstract}

\newpage

\section{Introduction}

There is mounting evidence that both the spectrum of anomalous
dimensions of infinitely long operators in 
${\cal N}=4$ super-Yang-Mills theory and the spectrum of the world
sheet sigma model (defined on a plane)  
can be described in terms of Bethe ${\rm ans{\hskip
-1.4pt}\ddot{\,a}tze}$. They are based on the scattering matrices of  
the fundamental excitations building, respectively, the gauge theory
gauge invariant operators and the  
physical string states.  The same information is encoded in the scattering 
matrix of the momentum eigenstate presentation of these excitations --
the magnon scattering matrix. 
There currently exists an all-loop conjecture for the scattering of
gauge theory magnons \cite{BDS_p, BeSt} 
as well as indirect results for the leading order \cite{AFS} and the first
subleading correction \cite{HeLo, fk} to the scattering phase of   
string theory magnons. They have been argued for and tested in detail in
the large 't~Hooft coupling  
and small world sheet momentum regime, in which the gauge theory
magnons are in one to one correspondence  
with the world sheet fields in the uniform gauge. The string theory
magnon scattering matrix is further  
conjectured to have the same expression even if the world sheet
momenta are held fixed in the large 't~Hooft  
coupling limit. It is important to subject it to controlled tests in this regime.

More generally, an algebraically-determined scattering matrix needs to
be subjected to consistency tests. 
A distinct possibility is that this S-matrix is unrelated to the
Lagrangian one would like to quantize. 
If the states scattered by it are directly related to the fields of
the original Lagrangian, the  
details of the S-matrix may be tested by direct perturbative higher
order calculation.  
\footnote{It is worth mentioning that naive ``exact'' S-matrices for the
non-simply laced affine Toda theories fail this test at the 1-loop level 
\cite{No_S_mat, Delius}. To a certain extent, the calculation mentioned  
here explicitly tests the quantum integrability of the theory. These theories are
nevertheless integrable at the quantum level. The ``true''
exact S-matrices for the non-simply laced affine Toda theories were constructed 
and analyzed in \cite{DGZa, CDSa, PDorey2}. Their consistency with perturbation 
theory is subtle and interesting. I would like to thank Patrick Dorey
for clarifying 
this to me.} 
If the exact S-matrix describes physical bound states one may, in the
same spirit, test whether their  
scattering is correctly reproduced by corrections to the classical
scattering induced by the original  
Lagrangian.

In an arbitrary field theory, the scattering of bound states (if they
exist) is related in a rather  
complicated way to the scattering of the fundamental fields. The
bootstrap approach was proposed as a  
way of determining both the spectrum of bound states and their
scattering amplitudes. Its main postulates  
are that 

1) the scattering amplitudes are determined self-consistently; all
particles that can appear as scattering  
states are the same particles being exchanged in the scattering
process and thus the S-matrix exhibits  
physical simple poles corresponding the their going
on-shell. \footnote{It is worth emphasizing that  
it is not necessary that any pole satisfying physical state conditions
should correspond to a bound  
state. They  may simply correspond to ``fundamental'' states which
have been missed. A classic example  
is the appearance of the closed string poles in the open string
scattering amplitudes.} 
 
2) the S-matrix is unitary (in a generalized sense) and has prescribed
analytic properties. 

\vspace{2pt}

\noindent
In two-dimensional relativistic unitary integrable quantum field
theories these postulates have been used to great effect to determine
exact S-matrices. The calculation of bound state scattering matrices
is simplified by the lack of particle production in that it is
determined (up to an overall phase) by that of the fundamental
excitations. The successful comparison of the bootstrap-constructed
bound state S-matrix and the semiclassical Lagrangian calculation of
the same quantity is a test of the consistency of the S-matrix of the
fundamental excitations.

In the context of the  AdS/CFT correspondence,  it has been shown in
\cite{HoMa} that, after relaxing the level matching condition, 
there exist classical solutions with finite momenta of the world sheet
theory whose semiclassical scattering  reproduces the large 't~Hooft 
coupling limit of the AFS scattering matrix (with fixed momentum). 
Solitons of higher charges (dyonic giant magnons) and in larger sectors
have been also constructed \cite{Dorey1, Dorey2, MTT3, SpVo,
KRT}. They carry two R-charges. One of them, $J_1$, is strictly
infinite \footnote{See \cite{AFZ} for a discussion of $1/J_1$
corrections.} leading to an infinitely-long string. The
second charge $J_2$ corresponds to rotation in the direction
orthogonal to $J_1$. 

The charges of semiclassical solitons are physically unrelated to the
 't~Hooft coupling. However,  since the classical sigma model
corresponds  to infinite 't~Hooft coupling and the second charge $J_2$
can be (in principle)  arbitrary, there are (at least) two natural
regimes  one may consider:  $i)$ $J_2$ is fixed in the large 't~Hooft
coupling limit or $ii)$ $J_2$  scales with the 't~Hooft coupling 
as suggested by the
classical Lagrangian: $J_2\propto\sqrt{\lambda}$. 
While both situations formally appear as classical solutions of the
sigma model, all solutions of the former type differ only by
parametrically small terms (suppressed by factors of
$1/\sqrt{\lambda}$) and thus it is not completely clear why they
should, by themselves, be considered as distinct and trustworthy
classical solutions. One way of understanding them is as a small
charge limit of solutions of the second type.  

Perhaps a more controlled set of states are those whose charges scale
as indicated by the classical Lagrangian $J\propto\sqrt{\lambda}$.  As
discussed in \cite{MTT3}, they are visible in the finite gap equations
describing the $SU(2)$ sector and, apart from exhibiting finite world
sheet momentum, appear in the same parameter space as the finite
density  configurations: 
\begin{eqnarray}
	E-L=\sqrt{J_2^2+4{\bar\lambda}\sin^2\frac{p}{2}}-J_2 
	\equiv\sqrt{\bar\lambda}f(J_2/\sqrt{\bar\lambda})
	~~~~~~~{\cal J}_2=J_2/\sqrt{\bar\lambda}={\rm fixed}~~~~~~
{\bar\lambda}=\frac{\lambda}{4\pi^2}
~~.
\end{eqnarray}
The ratio ${\cal J}_2=J_2/\sqrt{\bar\lambda}$ can be treated as a free
parameter and $L=J_1+J_2$ is the length of the string (and of the
corresponding gauge theory operator). It has been argued \cite{Dorey2}
that, in the strong coupling limit, both the states with scaling and
fixed charges (in the sense emphasized above) correspond to the
solitons of the complex sine-Gordon theory (CsG).  

In this note we will compare the scattering of semiclassical strings corresponding 
to magnon bound states as predicted by the conjectured string magnon scattering matrix
\begin{eqnarray}
S_{\rm string}=S_{\rm gauge}\sigma^2~~~~~~\sigma^2=e^{i\theta_{\rm AFS}}
\label{string_SM}
\end{eqnarray}
and by the classical sigma model S-matrix. Here $S_{\rm gauge}$ denotes the 
conjectured all-loop gauge theory magnon S-matrix \cite{BDS_p, BeSt}.
The``dressing phase'' $\sigma$ was originally constructed \cite{AFS} to leading
order in the large 't~Hooft coupling limit by analyzing states 
with two angular momenta $J_1$ and $J_2$ of the same order and with small world sheet 
momenta. In this regime there is no sharp separation between 
quantum (i.e. $1/\sqrt{\bar\lambda}$) and finite size (i.e. $1/L\sim 1/J$) 
corrections; $\sigma$ captures -- through the Bethe equations -- 
the analytic part of the one-loop corrections to semiclassical string states.
The regime we will be probing is quite different: the angular momentum $J_1$
is strictly infinite, the world sheet momenta is kept fixed and the second 
angular momentum is fixed in units of $\sqrt{{\bar\lambda}}$.
The fact that the ``dressing phase'' 
$\sigma$ was constructed  in a different regime than the one in which
the bound states appear makes our comparison nontrivial.  
Within our setup we will interpret the states with fixed charge 
as the $J_2/\sqrt{{\bar\lambda}}\rightarrow 0$ limit of
$J_2/\sqrt{{\bar\lambda}}={\rm fixed}$ states.

After a brief discussion of bound states in general integrable field theories 
we proceed in \S3 with a review of the construction of bound state
S-matrices via the fusion procedure. We will not assume
two-dimensional Lorentz invariance, having in mind applications to the
gauge and string side of the AdS/CFT correspondence for which the
relevant S-matrices do not exhibit this symmetry. While this
makes the original relativistic analysis inapplicable, a weaker set of
assumptions (covering both the gauge and string S-matrices) leads to a
general condition for the unitarity of the bound state
S-matrices. This also leads to a condition that the state
corresponding to a simple pole is physical (i.e. that its wave
function is normalizable). For certain choices of variables this
condition has a simple form. 

In \S4 we will then apply this procedure to the scattering of bound
states described by the BDS S-matrix as well as to the bound states
described by the general ansatz for the string scattering matrix. 
We will find that for finite charge states the
scattering phase receives two contributions of the same order -- one
comes from the dressing phase while the second one is generated by the
prefactor $S_{\rm gauge}$ -- and discuss whether both should be
visible in the world sheet sigma model. We
will also notice that the physical state conditions prevent the
existence of bound states in some sectors, in agreement with previous
studies \cite{MTT3}. In the weak coupling limit we will nevertheless
find the partners of the magnon bound states in other sectors by
analyzing the $SU(2|2)^2$-invariant S-matrix.

We will then proceed to reconstruct the scattering phase of the sigma
model solitons from that of their CsG counterparts originally
discussed in \cite{dev2, DoHo} and then to compare
the result with the prediction of the fusion construction. We will
find that they agree and that there are two ways to interpret this
agreement. On the one hand, the contribution of the dressing phase is
correctly reproduced by the soliton-{\it anti\hphantom{{$\,$}}}soliton scattering while
that of $S_{\rm gauge}$ is interpreted as a coherent superposition of
one-loop effects. On the other hand, the sum of the contribution of
the dressing phase and of $S_{\rm gauge}$ is correctly reproduced by
the soliton-soliton scattering. We independently confirm these
identifications in the small momentum limit in \S8.

\section{Bound states and 2-particle S-matrices}

The spectrum of bound states of a field theory typically has a complicated 
structure. In particular, there can be bound states of two or more
particles, perhaps forming various representations of some symmetry group. 
The fundamental property of an integrable field theory defined on a plane
is that all physical information is encoded in the 2-particle scattering matrix
for its fundamental excitations.\footnote{It is typically not {\it a priori} clear  
what are the fundamental excitation of a theory and what is their relation to the fields 
appearing in its Lagrangian. There exist examples in which the fields appearing in 
the Lagrangian are bound states of some such ``more fundamental''
excitations (Toda field theories). }  
In particular, all information about bound states of arbitrary charge is encoded in it. 
While it is relatively clear that two-particle bound states may appear as poles in 
the two-particle S-matrix, it is also intuitively clear that many-particle bound states 
cannot directly appear in the fundamental 2-particle S-matrix. Indeed, 
let us assume that the scattered particles carry a representation of some group and 
that their scattering matrix respects this symmetry (which may not be
the complete symmetry of the S-matrix). It follows then that the bound
states  that are visible in the two-particle S-matrix are those
transforming in (irreducible) representations appearing in the direct
product of the representations labeling this S-matrix.  

Rather, many-particle bound states are described by the poles of the many-particle S-matrix 
which, in turn is determined by the 2-particle one. In the context of the AdS/CFT 
correspondence, multi-magnon bound states and the corresponding world sheet solitons 
have been identified in \cite{Dorey1, MTT3} in compact sectors of the theory. They are 
the so-called Bethe strings. From the stand point of the gauge theory spin chain, they 
represent deformations of the Bethe strings of the Heisenberg chain. The details
of both the conjectured gauge and string theory S-matrices imply that, in fact, on the 
rapidity plane $u(p)$ 
the structure of the Bethe strings remains the same as for the Heisenberg 
chain -- two rapidities differ by an integer multiple of the imaginary 
unit. In momentum space their shape changes in a coupling constant dependent way.

While the identification of the Bethe strings in the multi-particle scattering matrix 
is a perfectly valid standpoint, it is somewhat complicated to extract and justify the properties of these 
bound states. For this purpose it 
is perhaps more useful to build bound states ``one magnon at a time''.

An interesting question is whether all possible bound states can be identified this way. 
One may expect that magnon bound states are only part of the total
number of bound states. Also, since the magnon scattering matrix is diagonal, it is not 
immediately clear how the symmetries are realized. We will see this
more explicitly in the following sections.
More generally, the identification of {\em all} the bound states of a theory 
depends on one's ability to identify the fundamental 
excitations and determine their scattering matrix. This is relatively clear by considering 
the manifest symmetry group of the scattering matrix. The fundamental excitations as well 
as their bound states form linear representations of this group. Then, up to accidental 
zeros of the S-matrix, the corresponding Clebsch-Gordan coefficients determine which bound 
states are formed. Consequently, to identify all possible bound states it is necessary to 
identify the excitations carrying the smallest possible representation.\footnote{For bosonic 
groups this is the fundamental representation. It is however important to emphasize that
for supergroups the smallest representation may in fact be unphysical. Nevertheless, the 
scattering of the excitations carrying this (unphysical) representation 
determines the S-matrices of the physical bound 
states. Alternatively, in the space of physical states it is necessary to have the fundamental 
excitations transforming in a reducible representation of the (perhaps nonlinearly-realized) 
symmetry group.}

\section{Bootstrap and Fusion \label{footstrap}}

A very successful application of the bootstrap and fusion ideas is in
the context of integrable field theories, where they have been
successfully used to derive the scattering matrices of bound
states. In certain cases the result was successfully justified by
other means. We will review here the fusion construction \cite{ZaZa, Karo,
OgRW} being however careful not to assume that the scattering matrix
is relativistically invariant and derive the conditions that bound
states are physical. 

Let us assume that for some values of rapidities 
\begin{eqnarray}
	u_1=a_1+i b_1~~~~~~~~u_2=a_2+ib_2
	\label{pole}
\end{eqnarray}
the fundamental S-matrix $S_{12}$ has a simple pole with residue $R_{12}$. 
The precise values of $(a_1, a_2, b_1, b_2)$ depend on the details of $S_{12}$.
This residue may be decomposed in projectors onto representations of 
the manifest symmetry group of the S-matrix:
\begin{eqnarray}
	R_{12}=\sum_a~R_{12}^aP_a~~.
	\label{decomposition}
\end{eqnarray}
It is worth pointing out that, if the S-matrix has a larger symmetry than 
that which is manifest, the various representations appearing in (\ref{decomposition})
should realize (perhaps nonlinearly) this larger symmetry group.

From here it is natural to infer that the scattering matrix of the $(12)$ bound state
with the particle $3$ is proportional to the residue of the pole at
the position (\ref{pole}) of $S_{123}=S_{12}S_{13}S_{23}$. In general
this residue does not satisfy the Yang-Baxter equation or unitarity. The ansatz for the
S-matrix is  
\begin{eqnarray}
S_{(12)3}=A\,{\rm Res}_{(12)}(S_{12}S_{13}S_{23})B=AR_{12}S_{{\hat 1}3}S_{{\hat 2}3}\,B
\label{ansatz}
\end{eqnarray}
with $A$ and $B$ determined by unitarity and factorization. The
hatted indices denote the fact that the corresponding 
spectral parameters are evaluated on (\ref{pole}). 

From the Yang-Baxter equation it is trivial to read that
\begin{eqnarray}
	R_{12}S_{{\hat 1}3}S_{{\hat 2}3}=S_{{\hat 2}3}S_{{\hat 1} 3}R_{12}
\end{eqnarray}
while projecting from the left and from the right this equation onto 
$\sum_a \! P_a$ and $1-\sum_a \! P_a$ we find
\begin{eqnarray}
	(1-\sum_a \! P_a)\, S_{{\hat 2}3}S_{{\hat 1}3}R_{12}\, \sum_b \! P_b=0~~.
\end{eqnarray}

Requiring that the Yang-Baxter equation is satisfied by the ansatz (\ref{ansatz})
implies that, up to an overall function, $A$ and $B$ must satisfy
\begin{eqnarray}
	BAR_{12}=\sum_a P_a~~.
	\label{factorization}
\end{eqnarray}
The calculation is the same as in a relativistic field theory. 

Relativistic invariance makes it easy to
extract general information from the requirement that the 
S-matrix (\ref{ansatz}) is unitary
\begin{eqnarray}
 S_{(12)3}^\dagger S_{(12)3}=1~~.
 \label{unitarity}
\end{eqnarray} 
We are however interested in more general 
situations, such as the S-matrices appearing on the gauge theory or on the 
string theory side of the AdS/CFT correspondence. It turns out that the various conjectured
S-matrices obey the following identity
\begin{eqnarray}
	S_{{\hat 1}3}(u_1,u_3)^\dagger=S_{{\hat 1}3}(u_3,u_{1}^*)=S_{{\hat 1}3}(u_3,u_2)=E_{12} S_{{\hat 2}3}(u_3,u_2)E_{12}
	=E_{12} S_{{\hat 2}3}(u_2,u_3)^{-1}E_{12}
\label{identity}
\end{eqnarray}
where $E_{12}$ is the operator switching the labels of states $1$ and $2$ while leaving 
their momenta unchanged:
\begin{eqnarray}
	E_{12}|\alpha(1)\beta(2)\rangle=|\beta(1)\alpha(2)\rangle~~.
\end{eqnarray}
We also used the fact that the bound state $(12)$ appears if the $u$-parameters of the states 
$1$ and $2$ are complex conjugates of each other. Using the identity (\ref{identity}) 
it is not hard to show that (\ref{unitarity}) is satisfied if 
\begin{eqnarray}
	E_{12}R_{12}^\dagger A^\dagger A={\cal C}\sum_a P_a~~,
	\label{unit_constraint}
\end{eqnarray}
where ${\cal C}$ is an arbitrary function. Besides determining $A$ this condition 
identifies which 
poles are physical and which are not. Consider acting with (\ref{unit_constraint})
on an eigenstate in a definite representation and let us denote 
by $\eta_a$ the eigenvalue of the operator $E_{12}$ corresponding to this state. 
The operator $A^\dagger A$ is positive definite. Using the spectral 
decomposition (\ref{decomposition}) of the residue $R_{12}$ it follows that
\begin{eqnarray}
	{\rm sgn}({\cal C})\,\eta_a\,R_{12}^a>0~~.
	\label{physical_state}
\end{eqnarray}
There is still an ambiguity due to the unknown sign of ${\cal C}$. However, the 
important point is that the same function ${\cal C}$ appears for all states and thus 
we may determine its sign from one state. For relativistic field theories this translates
into \cite{Karo, OgRW}
\begin{eqnarray}
\eta_a\,R_{12}^a<0~~.
\end{eqnarray}
For nonrelativistic scattering matrices in general and for the BDS S-matrix in particular 
it is less clear how to construct a simplified form of the physical
state condition. Assuming that the pole of the magnon scattering matrix indeed
describes a physical 2-magnon bound state,  it appears possible to rephrase
(\ref{physical_state}) on the rapidity plane $u$ as the condition that
the imaginary part of $R_{12}^a$ has  a definite sign. 
This can be simply stated if we notice that, even though the relevant
S-matrices depend separately on the rapidities of the scattered
excitations,  the pole occurs for a fixed value for their difference. 
Then, the physical bound state condition (\ref{physical_state})  suggests that 
the imaginary part of the residues (taken with respect to
the variable in which the pole occurs at a {\em positive} multiple of the
imaginary unit) corresponding to physical bound states are positive   
\begin{eqnarray}
\Im(R_{12}^a)>0~~.
\label{phys_nr}
\end{eqnarray}
This is the condition we will use in the following. 
It is important to stress however that on the momentum plane the physical state condition 
appears to necessarily involve the phase $\eta$ in a nontrivial way.

Solving the conditions (\ref{factorization}) and (\ref{unit_constraint}) leads to the conclusion
that the scattering matrix of a bound state against a fundamental excitation is given by:
\begin{eqnarray}
S_{(12)3}=\sum_a (R_{12}^a)^{1/2}P_a~
(S_{13}S_{23})\Big|_{\rm physical~pole \/}\,\sum_a (R_{12}^a)^{-1/2}P_a~~.
\label{final_fusion}
\end{eqnarray}
While derived here from weaker assumptions, the expression for the
scattering matrix of a bound state off an elementary excitation is formally
as in relativistically-invariant theories
\cite{Karo, OgRW}.

The algorithm described here may be used to construct the scattering
matrix of more complex 
bound states (i.e. bound states of more than one excitation) off  elementary excitations 
as well as the scattering matrix of  bound states against each other.
For rank one S-matrices this expression simplifies 
considerably. In particular, there is no projection operator that is needed and moreover 
the residues of the poles cancel out and therefore are not needed (beyond making sure 
that the corresponding pole is physical).

Quite clearly, the fusion algorithm applies to a large class of S-matrices; in particular, 
it is not restricted to unit rank. However, unit rank sectors of
larger S-matrices are particularly simple to analyze. As mentioned
before, the spectrum of bound states obtained in such sectors
naturally extends to representations of the manifest symmetry group of
the original S-matrix.  If the complete S-matrix is more symmetric,
then these representations should in turn fit into representations of
this larger symmetry group. It is in general unclear how this happens
or whether the symmetry is linearly realized at the level of the bound
states. In the following we will not address this issue and we will
mostly restrict ourselves to unit rank sectors of the gauge and string theory
S-matrix. We will however identify (in the small 't~Hooft coupling
limit) the S-matrix whose poles corresponds to the multiplet
containing the 2-magnon bound state and construct the S-matrix of this
multiplet off the ``elementary'' excitations.

\section{BDS and AFS type S-matrices and the scattering of bound
states \label{various_S_matrices}} 

We will now apply the algorithm described in the previous section and
determine the scattering matrix of multi-magnon bound states as
encoded both in the BDS \cite{BDS_p} and in the AFS \cite{AFS} 
S-matrices. We will then take
the strong coupling limit on the results obtained from the AFS
S-matrix and compare it with the soliton scattering in the sigma
model. 

The part of the $SU(2)$ sector S-matrix carrying information about the bound states is 
\begin{eqnarray}
S_{12}^{\rm BDS}=\frac{u(p_1)-u(p_2)+i}{u(p_1)-u(p_2)-i}~~~~~~~~~
u(p)=\frac{1}{2}\cot\frac{p}{2}
\sqrt{1+4{\bar\lambda}\sin^2\frac{p}{2}}~~.
\label{BDS}
\end{eqnarray}
Up to the precise form of $u(p)$, this S-matrix is the same as that
describing magnon scattering in the Heisenberg chain. The different
$u(p)$ has some nontrivial effects in that the actual physical
parameter is the momentum rather than $u$ and the bound states
correspond to some complex values of $p$. It is nevertheless possible
to figure out the real part of $u(p)$ corresponding to bound states:
in particular, for a 2-magnon bound state it turns out \cite{Dorey1,MTT3}
that 
\begin{eqnarray}
u_2(p)=\frac{1}{2}\cot\frac{p}{2}\sqrt{4+4{\bar\lambda}\sin^2\frac{p}{2}}~~.
\label{2bound}
\end{eqnarray}
A similar expression for a $J$-magnon bound state was obtained in
\cite{MTT3} starting from the multi-magnon S-matrix (Bethe equations)
\begin{eqnarray}
u_J(p) = \frac{1}{2}\cot\frac{p}{2}\sqrt{J^2+4{\bar\lambda}\sin^2\frac{p}{2}}~~.
	\label{bound_state_rapidity}
\end{eqnarray}
Here $p$ denotes the total momentum of the bound state.

With this input it is easy to proceed with fusing (\ref{BDS}) into the
scattering of bound states. 

\subsection{BDS}

We have used the pole in the BDS S-matrix
(\ref{BDS}) at $u(p_2)-u(p_1)=i$ as input in our simplified form of the
physical state condition.  It is not hard to see that $\Im({\rm Res}S_{12}^{\rm
BDS})=2$. It follows then that the scattering matrix of the charge$-2$ bound
state against an elementary magnon is  
\begin{eqnarray}
S_{J_2=2, 1}=\frac{u(p_2)-u_1+i}{u(p_2)-u_1-i}\,\frac{u(p_3)-u_1+i}{u(p_3)-u_1-i}
\Big|_{{u(p_2)=u_{{J_2=2}}+i/2 \atop u(p_3)=u_{J_2=2}-i/2}} =
\frac{u_{J_2=2}-u_1+{\textstyle{\frac{3i}{2}}}}{u_{J_2=2}-u_1-{\textstyle{\frac{3i}{2}}}}
	\,
\frac{u_{J_2=2}-u_1+{\textstyle{\frac{i}{2}}}}{u_{J_2=2}-u_1-{\textstyle{\frac{i}{2}}}}~~.
\end{eqnarray}
Out of the two poles the physical one is at
$u_{J_2=2}-u_1={\textstyle{\frac{3i}{2}}}$. The imaginary part of the
residue of the other pole (in the sense described above) is 
\begin{eqnarray}
\Im\left({\rm Res}S_{J_2=2, 1}\Big|_{u_{J_2=2}-u_1={\textstyle{\frac{i}{2}}}}\right)=-2
\end{eqnarray}
implying that this second pole is unphysical.

Repeating $J_2$ times the fusion procedure for the physical pole it is
quite easy to find that the scattering matrix of a $J_2$-magnon bound
state against an elementary magnon state is 
\begin{eqnarray}
S_{J_2,1}=\frac{u_{J_2}-u_1+\frac{i}{2}(J_2+1)}{u_{J_2}-u_1-\frac{i}{2}(J_2+1)}
\frac{u_{J_2}-u_1+\frac{i}{2}(J_2-1)}{u_{J_2}-u_1-\frac{i}{2}(J_2-1)}~~.
\label{Jon1}
\end{eqnarray}
The only physical pole of this S-matrix -- which is the pole that
should be used to raise $J_2$ by one unit -- is at
\begin{eqnarray}
u_{J_2}-u_1=\frac{i}{2}(J_2+1)~~.
\end{eqnarray}
Here, $u_J$ continues to carry the interpretation of rapidity/spectral
parameter but, as mentioned before, its expression in terms of the
bound state momentum is different from that of elementary magnons and
is given by (\ref{bound_state_rapidity}). 

We may further fuse the S-matrices (\ref{Jon1}) into those describing the scattering of a 
$J_2^{(1)}$-magnon bound state against a $J_2^{(2)}$-magnon bound
state with $J_2^{(1)}>J_2^{(2)}$. The result is: 
\begin{eqnarray}
\label{BDS_totally_fused}
&&	S_{J_2^{(1)}{J_2^{(2)}}}^{\rm BDS}=\frac{u_{J_2^{(1)}}-u_{J_2^{(2)}}+
\frac{i}{2}(J_2^{(1)}-{J_2^{(2)}})}{u_{J_2^{(1)}}-u_{J_2^{(2)}}-\frac{i}{2}(J_2^{(1)}-{J_2^{(2)}})}
\times\\
&&~~~~~~~~~~~~~~~~~~
\times	\left[\prod_{l=1}^{{J_2^{(2)}}-1}  	
\frac{u_{J_2^{(1)}}-u_{J_2^{(2)}}+\frac{i}{2}(J_2^{(1)}-{J_2^{(2)}}+2l)}
{u_{J_2^{(1)}}-u_{J_2^{(2)}}-\frac{i}{2}(J_2^{(1)}-{J_2^{(2)}}+2l)}
\right]^2 
\frac{u_{J_2^{(1)}}-u_{J_2^{(2)}}+\frac{i}{2}(J_2^{(1)}+{J_2^{(2)}})}
{u_{J_2^{(1)}}-u_{J_2^{(2)}}-\frac{i}{2}(J_2^{(1)}+{J_2^{(2)}})}~~.
\nonumber
\end{eqnarray}
In this case both $u_{J_2^{(1)}}(p)$ and $u_{J_2^{(2)}}(p)$  are given
by (\ref{bound_state_rapidity}). 

An interesting feature of (\ref{BDS_totally_fused}) is that, besides
simple poles, it also exhibits higher  order poles. This structure was
previously observed in integrable relativistic and nonrelativistic
field theories \cite{Yang, Coleman, Braden, Christe, Delius}. In
relativistic theories they were
associated to anomalous thresholds in the 
higher loop contributions to the scattering matrix of
bound states. It is not currently clear whether the double-poles we
find here have an interpretation from the standpoint of the
nonrelativistic field theory whose resumed scattering matrix
reproduces (\ref{BDS}) or whether they can be reproduced only in 
the complete world sheet theory.

\subsection{AFS}

The results obtained above can be easily extended to include for the conjectured 
dressing phase connecting the gauge theory and the string theory S-matrices.
\begin{eqnarray}
S_{\rm string}=\,S_0\,S_{\rm su(2|2)^2}
~~~~
\rightarrow
~~~~
S_{\rm AFS}=\sigma^2 S_{\rm BDS}~~.
\label{string_S}
\end{eqnarray}
In uniform gauge the general structure of $\sigma^2$ is believed to have the form
\begin{eqnarray}
\sigma^2(p_1,p_2)\equiv e^{i\theta }
=\exp\left\{{i\sum_{rs}\,
\left(\frac{{\bar\lambda}}{4}\right)^{\frac{1}{2}(r+s-1)}\,c_{rs}(\sqrt{\bar\lambda})\,
\Big(q_r(p_1)q_s(p_2)-q_s(p_1)q_r(p_2)\Big)}\right\}
\label{phase}
\end{eqnarray}
where $q_r$ are the higher conserved local charges. The expression of
the coefficients $c_{rs}(\sqrt{\bar\lambda})$ is not known, but there
exists indirect evidence that the first few terms in their expansion
are (\cite{AFS} and \cite{HeLo, fk}, respectively) 
\begin{eqnarray}
	c_{rs}(\sqrt{\bar\lambda})=\delta_{s,r+1}+\frac{(-)^{r+s}-1}{2\sqrt{\bar\lambda}}\,
	\frac{(r-1)(s-1)}{(r-1)^2-(s-1)^2}+\dots~~.
\end{eqnarray}

It is not hard to see using the resummed expression for (\ref{phase})
given in \cite{BeSt, BeTs} that the leading term in the dressing phase
does not introduce additional poles in the 2-particle S-matrix 
besides those already present in $S_{\rm BDS}$. 
It therefore follows that the contribution of the
residues of the physical pole to the bound state $S$ matrix cancels
out.\footnote{This is valid in general if the incoming and outgoing
particles belong to {\it the same} rank one sector.} Thus, the
remaining contribution comes from multiplying together the dressing
factors and evaluating them for the complex momenta describing the
bound state formation. The result emulates the elementary magnon
dressing phase (\ref{phase}) except that the higher local charges now
are those of the corresponding bound states.  
\begin{eqnarray}
	q_r^J(p_{\rm total})=\sum_{l=-J}^{J}q_r(a_l+ib_l)~~.
\end{eqnarray}
This expression can be substantially simplified by recalling that
$q_r(p)$ may be written in terms of the rapidities $u(p)$. In these
variables the charges become 
\begin{eqnarray}
q_r^J=\frac{i}{r-1}\left[\frac{1}{x(u_J(p)+{\textstyle{\frac{i}{2}}}J)^{r-1}}
-\frac{1}{x(u_J(p)-{\textstyle{\frac{i}{2}}}J)^{r-1}}\right]~~~~~~{\rm
with}~~~~x(u)=\frac{1}{2}(u+\sqrt{u^2-{\bar\lambda}})~~. 
\end{eqnarray}
Following our discussion in \S\ref{footstrap} as well as that in the
previous subsection we will treat the rapidity $u$ rather than $x(u)$ as
fundamental variable. The appearance of $x(u)$ should be thought of
as a shorthand for its expression in terms of the rapidity (or
momentum). 

With this clarifications, the conjectured general form of the string
S-matrix (\ref{string_S}) implies that the scattering matrix of a
$J_2^{(1)}$- and a ${J_2^{(2)}}$-magnon bound state is: 
\begin{eqnarray}
S_{J_2^{(1)}{J_2^{(2)}}}&=&
\frac{u_{J_2^{(1)}}-u_{J_2^{(2)}}+\frac{i}{2}(J_2^{(1)}+{J_2^{(2)}})}{u_{J_2^{(1)}}-u_{J_2^{(2)}}
-\frac{i}{2}(J_2^{(1)}+{J_2^{(2)}})}\times\\
&&~~~~
\times	\left[\prod_{l=1}^{J_2^{(2)}-1}  	
\frac{u_{J_2^{(1)}}-u_{J_2^{(2)}}+\frac{i}{2}(J_2^{(1)}-J_2^{(2)}+2l)}
{u_{J_2^{(1)}}-u_{J_2^{(2)}}-\frac{i}{2}(J_2^{(1)}-J_2^{(2)}+2l)} \right]^2
\frac{u_{J_2^{(1)}}-u_{J_2^{(2)}}+\frac{i}{2}(J_2^{(1)}-{J_2^{(2)}})}{u_{J_2^{(1)}}-u_{J_2^{(2)}}
-\frac{i}{2}(J_2^{(1)}-{J_2^{(2)}})}
\times \cr
&&~~~~	
\times
\exp\left\{i\sum_{rs}c_{rs}(\sqrt{\bar\lambda})
\left(\frac{\bar\lambda}{4}\right)^{\frac{1}{2}(r+s-1)}
\Big(q_r^{J_2^{(1)}}q_s^{J_2^{(2)}}-q_s^{J_2^{(1)}}q_r^{J_2^{(2)}}\Big)\right\}~~.
\end{eqnarray}
The self-similar property of the dressing phase is a direct
consequence of its bilinear dependence on the higher local charges. 

\section{The large $\lambda$ limit}

For the purpose of comparison with the world sheet theory we must take the large 
't~Hooft coupling limit while keeping all momenta fixed. In this limit
only the leading term in the coefficients $c_{rs}(\sqrt{\bar\lambda})$
is necessary. As emphasized in the introduction, there are in fact
several  distinct $\lambda\rightarrow\infty$ limits, which are
distinguished by whether the length of the Bethe string is kept finite
or is allowed to scale to infinity. Following our philosophy, we
express the relevant  higher local charges and the corresponding
spectral parameters $x$ in terms of the rapidities $u$ and in terms of
the bound state momenta: 
\begin{eqnarray}
\label{full_charges}
x&=&\frac{1}{2}\left(u+\sqrt{u^2-{\bar\lambda}}\right) \\
x^{\pm(J)}(p)&\equiv&x(u_J(p)\pm {\textstyle{\frac{i}{2}}}J)
=\frac{e^{\pm\frac{i}{2}p}}{ 4\sin\frac{p}{2}}
\left(J+\sqrt{J^2+4{\bar\lambda}\sin^2\frac{p}{2}}\right)\cr
q_r^J(p)&=&\frac{i}{r-1}\left(\frac{1}{x^{+(J)}(p)^{r-1}}-\frac{1}{x^{-(J)}(p)^{r-1}}\right)=
\frac{2}{r-1}\sin\textstyle{\frac{1}{2}}(r-1)p
\left[\frac{\sqrt{J^2+4{\bar\lambda}
\sin^2\frac{p}{2}}-J}{{\bar\lambda}\sin\frac{p}{2}}\right]^{r-1}~~.
\nonumber
\end{eqnarray}
These expressions hold for any nonvanishing $J$. At the same time,
they define a prescription for taking the ${\cal J}_2^{(i)}\rightarrow
0$ limit or, alternatively, the large 't~Hooft coupling limit. Indeed,
from the standpoint of the spectral parameter of shifted argument $x(u_J\pm
iJ/2)$, the limit ${\cal J}=J/\sqrt{\bar\lambda}\rightarrow 0$ makes
$x^{\pm(J)}$ equal to leading  order. However, when the limit is taken
on their momentum space expressions (\ref{full_charges}) 
they remain different (cf. the second equation
above, where the momentum dependent phase survives the limit ${\cal
J}=J/\sqrt{\bar\lambda}\rightarrow 0$.). 

\

Let us first consider the case in which the length of the Bethe
strings is kept fixed in the sigma model limit. It is then easy to see
that all the local charges of the bound states become
equal to those of the elementary magnons.\footnote{The same result for
the energy is evident from \cite{Dorey1, Dorey2, MTT3}
\begin{eqnarray}
E_J=\sqrt{J^2+4{\bar\lambda}\sin^2\textstyle{\frac{1}{2}} p }-J
~~~~\stackrel{{\scriptstyle{\bar\lambda}\rightarrow\infty}}{\lin\!\!\!\!
\longrightarrow}
~~~~E_J=2\sqrt{{\bar\lambda}}\Big|
\sin\textstyle{\frac{1}{2}} p \Big|~~,
\nonumber
\end{eqnarray}
for any charge $J$.} Thus, in this limit the $J$-magnon bound states
are indistinguishable from the elementary magnons, suggesting that they
are distinguished only by quantum effects making suspicious a direct
relation between these states and classical string solutions. A
possible interpretation, which is the one we will adopt in the
following, is that they should be considered as corresponding to the
small charge limit of classical solutions. Proceeding along these
lines, it is relatively easy to see that the sigma model limit of
the fixed charge bound state scattering matrix is the same as that of
``elementary'' (i.e. unit charge) magnons 
\begin{eqnarray}
\theta_{0}=\frac{\sqrt{\bar\lambda}}{2\pi}\left(\cos p_{J_2^{(2)}}-\cos p_{J_2^{(1)}}\right)
\ln\frac{1-\cos\textstyle{\frac{1}{2}}(p_{J_2^{(1)}}-p_{J_2^{(2)}})}
{1-\cos\textstyle{\frac{1}{2}}(p_{J_2^{(1)}}+p_{J_2^{(2)}})}~~,
\label{HM_phase}
\end{eqnarray}
where $p_{J_2^{(1)}}$ and $p_{J_2^{(2)}}$ are the momenta of the
charge-${J_2^{(1)}}$ and charge-${J_2^{(2)}}$ magnons, respectively.  

\

The other -- more interesting -- limit is when the lengths of the
Bethe strings $J_2^{(i)}$ are scaled to infinity together with the
't~Hooft coupling such that ${\cal
J}_2^{(i)}=J_2^{(i)}/\sqrt{\bar\lambda}$ is kept fixed. This limit is
analogous to that isolating the classical states in the $SU(2)$ sector
and, by analogy with the states with nonzero filling fraction, should
be considered as corresponding to classical sigma model solutions.  

Since the expressions for the higher local charges are different for
the two scattered states, it is necessary to recompute the sum
(\ref{full_charges}) while taking this into account. It is
straightforward to do so by making use of the identity 
\begin{eqnarray}
\sum_{r\ge
2}\frac{1}{(r-1)r}\frac{{\bar\lambda}^{r-1/2}}{x^{r-1}y^r}=-\frac{1}{y\sqrt{\bar\lambda}}
-\frac{xy/{\bar\lambda}-1}{y/\sqrt{\bar\lambda}}\ln
\left(1-\frac{{\bar\lambda}}{xy}\right)~~. 
\end{eqnarray}
A small amount of algebra leads to\footnote{All the sums can be
performed using the momentum representation of the higher local
charges (\ref{full_charges}).} 
\begin{eqnarray}
\label{AFS_full}
\theta_{0}^{{J_2^{(1)}}{J_2^{(2)}}}&=&2(u_{J_2^{(2)}}-u_{J_2^{(1)}})
\ln\left[
\frac{1-\frac{{\bar\lambda}/4}{
x^{+({J_2^{(1)}})}x^{-({J_2^{(2)}})}}}{1-\frac{{\bar\lambda}/4}{
x^{+({J_2^{(1)}})}x^{+({J_2^{(2)}})}}} 
\frac{1-\frac{{\bar\lambda}/4}{
x^{-({J_2^{(1)}})}x^{+({J_2^{(2)}})}}}{1-\frac{{\bar\lambda}/4}{
x^{-({J_2^{(1)}})}x^{-({J_2^{(2)}})}}}\right]\\ 
&+&
i({J_2^{(1)}}+{J_2^{(2)}})\ln\left[
\frac{1-\frac{{\bar\lambda}/4}{
x^{-({J_2^{(1)}})}x^{+({J_2^{(2)}})}}}{1-\frac{{\bar\lambda}/4}{
x^{+({J_2^{(1)}})}x^{-({J_2^{(2)}})}}}\right]+ 
i({J_2^{(1)}}-{J_2^{(2)}})\ln\left[
\frac{1-\frac{{\bar\lambda}/4}{
x^{+({J_2^{(1)}})}x^{+({J_2^{(2)}})}}}{1-\frac{{\bar\lambda}/4}{
x^{-({J_2^{(1)}})}x^{-({J_2^{(2)}})}}}\right]~~. 
\nonumber
\end{eqnarray}
Clearly, $\theta_{0}^{{J_2^{(1)}}{J_2^{(2)}}}$ reduces to $\theta_{\rm
AFS}$ of \cite{BeTs,BeSt} for $J_2^{(1)}=J_2^{(2)}=1$, as it
should. It is also trivial to recover the phase
(\ref{HM_phase}). Indeed, by keeping both ${J_2^{(1)}}$ and
${J_2^{(2)}}$ fixed while taking the large $\lambda$ limit it is easy
to see that the second line is subleading while the first line
immediately leads to (\ref{HM_phase}).  

If either ${J_2^{(1)}}$ or ${J_2^{(2)}}$ or both scale as
$\sqrt{\lambda}$ the second line in equation (\ref{AFS_full}) is no
longer subleading, being of the same order as the first line. It is in
fact possible to scale away the 't~Hooft coupling and express the
scattering matrix in terms of the ratios ${\cal
J}^{(i)}={J}^{(i)}/\sqrt{\bar\lambda}$. 

An interesting observation is that, if ${J_2^{(2)}}/\sqrt{\lambda}$ is
fixed in the large $\sqrt{\lambda}$ limit, there appears to be an
additional {\em leading order} contribution to the magnon scattering
phase: there are ${J_2^{(2)}}$ factors in $S^{\rm
BDS}_{J_2^{(1)}J_2^{(2)}}$ and thus its logarithm scales to infinity
as $\sqrt{\lambda}$. The potential shift of $\theta_0^{J_2^{(1)}J_2^{(2)}}$ is
\begin{eqnarray}
\label{potential_extra}
&&
\delta \theta_0^{J_2^{(1)}J_2^{(2)}}=i\ln\left\{
\frac{u_{J_2^{(1)}}-u_{J_2^{(2)}}-\frac{i}{2}({J_2^{(1)}}-{J_2^{(2)}})}
 {u_{J_2^{(1)}}-u_{J_2^{(2)}}+\frac{i}{2}({J_2^{(1)}}-{J_2^{(2)}})}\times\right.\\
&&~~~~~~~~
\times \left.	\left[
\prod_{l=1}^{{J_2^{(2)}}-1}  	
\frac{u_{J_2^{(1)}}-u_{J_2^{(2)}}-\frac{i}{2}({J_2^{(1)}}-{J_2^{(2)}}+2l)}
   {u_{J_2^{(1)}}-u_{J_2^{(2)}}+\frac{i}{2}({J_2^{(1)}}-{J_2^{(2)}}+2l)} \right]^2
\frac{u_{J_2^{(1)}}-u_{J_2^{(2)}}-\frac{i}{2}({J_2^{(1)}}+{J_2^{(2)}})}
   {u_{J_2^{(1)}}-u_{J_2^{(2)}}+\frac{i}{2}({J_2^{(1)}}+{J_2^{(2)}})}
\right\}~~.
\nonumber
\end{eqnarray}

There are two possible standpoints regarding the interpretation of this correction to 
the dressing phase contribution to the scattering phase of magnon bound states: 

1) $\delta \theta_0^{J_2^{(1)}J_2^{(2)}}$ contains $J_2^{(2)}$ terms and thus, 
under our assumptions, is of the same order as the contribution of the dressing phase. 
It is therefore tempting to expect that it is also visible in the classical sigma model. 

2) $\delta \theta_0^{J_2^{(1)}J_2^{(2)}}$ is not proportional to $\sqrt{{\bar\lambda}}$ 
(though scales like it under our assumptions) 
and thus it should not be included in a comparison with semiclassical string calculations 
which -- by construction -- yield contributions proportional to $\sqrt{\bar\lambda}$
to all quantities. From this standpoint the $\sqrt{{\bar\lambda}}$ dependence 
of (\ref{potential_extra}) suggests that $\delta \theta_0^{J_2^{(1)}J_2^{(2)}}$ has -- from 
the sigma model perspective -- a 1-loop origin and the apparent semiclassical scaling is 
due to a ``coherent superposition'' of quantum effects.

\vspace{2pt}

In \S\ref{comparison} we will see that both standpoints have an
interpretation in the world sheet sigma model. To analyze this issue
we will need a more tractable form for $\delta
\theta_0^{J_2^{(1)}J_2^{(2)}}$. Up to corrections of the order of
$1/J_2^{(i)}$ this can be done by replacing the sum of logarithms  
by their integral.\footnote{ The calculation is formally identical had we chosen to 
express it as a large $\lambda$ limit. This however obscures the fact
that the reason this replacement is reliable is that the charges
$J_2^{(i)}$ are large.} The result is: 
\begin{eqnarray}
\delta \theta_0^{J_2^{(1)}J_2^{(2)}}&=&~ 2(u_{J_2^{(2)}}-u_{J_2^{(1)}})
\ln\left[\frac{(u_{J_2^{(1)}}-u_{J_2^{(2)}})^2+{\textstyle{\frac{1}{4}}}(J_2^{(1)}-J_2^{(2)})^2}
{(u_{J_2^{(1)}}-u_{J_2^{(2)}})^2+{\textstyle{\frac{1}{4}}}(J_2^{(1)}+J_2^{(2)})^2}\right]\\
&&+i(J_2^{(1)}+J_2^{(2)})
\ln\left[\frac{u_{J_2^{(1)}}-u_{J_2^{(2)}}+{\textstyle{\frac{i}{2}}}(J_2^{(1)}+J_2^{(2)})}
{u_{J_2^{(1)}}-u_{J_2^{(2)}}-{\textstyle{\frac{i}{2}}}(J_2^{(1)}+J_2^{(2)})}\right]\cr
&&-i(J_2^{(1)}-J_2^{(2)})
\ln\left[\frac{u_{J_2^{(1)}}-u_{J_2^{(2)}}+{\textstyle{\frac{i}{2}}}(J_2^{(1)}-J_2^{(2)})}
{u_{J_2^{(1)}}-u_{J_2^{(2)}}-{\textstyle{\frac{i}{2}}}(J_2^{(1)}-J_2^{(2)})}\right]~~.
\nonumber
\end{eqnarray}

It is worth mentioning that an intermediate regime between the two limits discussed 
above consists of keeping one charge fixed while scaling the other one
to infinity together with $\lambda$. Since in our use of the fusion
rules we assumed that ${J_2^{(1)}}>{J_2^{(2)}}$, it is natural that we
keep ${J_2^{(2)}}$ fixed. We will not consider separately this regime,
as it may be trivially obtained as the
${J_2^{(2)}}/\sqrt{\bar\lambda}\rightarrow 0$ of the previous
regime. Similarly, if we take the limit
${J_2^{(i)}}/\sqrt{\bar\lambda}\rightarrow 0$ for both $i=1$ and $i=2$
we recover the first limit (\ref{HM_phase}). Assuming that the general
limit agrees with the sigma model scattering, this observation
justifies the interpretation of the fixed charge solitons
as the small charge limit of solitons present in the $SU(2)$
sector. This is similar to interpreting BMN states as small charge
limits of classical solutions of the sigma model. 


\subsection{On other sectors}

Let us briefly comment on the other rank one sectors
-- $SU(1|1)$ and $SL(2)$ -- as well as on the full S-matrix. 
Using the fact that the dressing  phase does not -- at least to this
order -- contain poles describing propagating modes, it suffices to analyze  
\begin{eqnarray}
S_\eta=\left(\frac{x^+_1-x^-_2}{x^-_1-x^+_2}\right)^\eta
\frac{1-\frac{\bar\lambda}{4 x^+_1x^-_2}}{1-\frac{\bar\lambda}{4 x^-_1x^+_2}}
\end{eqnarray}
for $\eta=0$ and $\eta=-1$.

It is relatively easy to search for poles in these scattering matrices
on the rapidity plane $u$. The result is that for $\eta=0$ there are
no poles while for $\eta=-1$ there is a pole which does not satisfy
the physical state condition. Thus, there is no bound state of two
magnons in the $SU(1|1)$ and the $SL(2)$ sectors. This latter
observation matches the conclusion of \cite{MTT3} that the only bound
states in the $SL(2)$ sector contain infinitely many magnons. 

Naively these observations present a puzzle. 
While symmetry considerations suggest that there should exist bound 
states in other sectors, the absence of bound states in other unit rank 
sectors may suggest that the bound states found in the $SU(2)$ sector 
do not have a supersymmetric completion.
This is however not true and the ``missing'' bound states may be identified 
as poles in the $SU(2|2)^2$-invariant S-matrix \cite{Besu22} rather than 
the magnon S-matrix.  Perhaps the easiest way to see this is to take the 
small coupling limit where the $SU(2|2)$ symmetry becomes manifest
\begin{eqnarray}
S_{\rm SU(2|2)}=
\frac{1}{2}E_{12}\,S_0\,\left[(1-{\cal P}_{12})
+\frac{u_2-u_1+i}{u_2-u_1-i}(1+{\cal P}_{12})\right]
\end{eqnarray}
where $E_{12}$ is the operator interchanging the rapidities of the excitations 
introduced in equation (\ref{identity}) (or, alternatively, switches the labels 
of the two excitations while leaving the momenta unchanged).  Also, ${\cal P}$ is the graded permutation 
operator and thus $(1\pm {\cal P}_{12})$ are projectors  onto the graded symmetric and graded
antisymmetric representations.  

Thus, at vanishing 't~Hooft coupling, $S_{\rm SU(2|2)}$ contain a
physical pole at $u_2-u_1=i$ whose residue is proportional to the
projector  onto the two-index graded symmetric representation of 
$SU(2|2)$. The vanishing 't~Hooft coupling limit of the ${\cal N}=4$ 
S-matrix \cite{Besu22} 
\begin{eqnarray}
S_{{\cal N}=4}=S_{\rm SU(2|2)}\otimes S_{\rm SU(2|2)} \frac{u_2-u_1-i}{u_2-u_1+i}
\end{eqnarray}
inherits this pole. Its corresponding residue is
\begin{eqnarray}
{\rm Res}S_{{\cal N}=4}\Big|_{u_2-u_1=i}=2i\,(1+{\cal P}_{12}) \otimes (1+{\cal P}_{12})
\end{eqnarray}
and projects onto the tensor product of two  2-index graded
symmetric representation of $SU(2|2)$. One of the states in this
representation is the bound state visible in the magnon
S-matrix. 

Thus, at vanishing  't~Hooft coupling, the $SU(2|2)^2$-symmetric
S-matrix contains sufficiently many poles to realize a representations
of the ${\cal N}=4$ symmetry algebra and the apparent contradiction
outlined in the beginning of this section is resolved in this limit.
While this analysis obviously does not directly apply at the level of
the sigma model, it is nevertheless interesting to note that it does
not follow the standard pattern of bosonic sigma models. There the
(CDD) dressing factor implies that bound states occur in antisymmetric
representations of the symmetry group and one may have expected here 
graded-antisymmetric representations to occur.  

Using the fusion algorithm it is easy to construct the scattering
matrix of bound states and elementary excitations. The calculation is
simplified by the fact that the scattering of two $SU(2|2)$
excitations produces a single physical bound state and thus, as for
unit rank sectors, the precise value of the residue is not relevant.
\begin{eqnarray}
S_{1b}=E_{b1}S_0(u_b+i/2, u_1)S_0(u_b-i/2, u_1)
\left[P_{\{2, 1, 0,\dots\}}+
\frac{u_b-u_1+{\textstyle{\frac{3i}{2}}}}{u_{b}-u_1-{\textstyle{\frac{3i}{2}}}}
	\,
\frac{u_b-u_1+{\textstyle{\frac{i}{2}}}}{u_b-u_1-{\textstyle{\frac{i}{2}}}}
\,P_{\{3,0,0,\dots\}}
\right]
\label{susy_bound}
\end{eqnarray}
where $P_{\{j_1,j_2,j_3,\dots\}}$ denotes the projector onto the
representation whose super-Young tableau has $j_k$ boxes on the $k$-th
row and $u_b$ denotes the rapidity of the 2-particle bound state
(\ref{2bound}). Similarly to the representation of the 
2-particle bound state, the structure of the S-matrix above 
is again different from that in the usual bosonic sigma models.  
Similarly to the $SU(2)$-magnon, the S-matrix (\ref{susy_bound}) has 
one unphysical and one physical pole, the latter of which may be used
to construct further bound states and their corresponding scattering matrices.

The situation is less clear at $\lambda\ne 0$, but it appears unlikely
that the representation of the bound state changes as a function of
the coupling constant.  \footnote{It is notable however that 
the imaginary parts of the residues of some of the elements of the
$SU(2|2)$-invariant S-matrix change  sign at rapidity-dependent
values of the 't~Hooft coupling. 
This should not be interpreted as a change in the 
spectrum of the theory. Inspecting the relevant matrix elements 
leads to the conclusion that the poles with indefinite-sign residue
arise in elements which vanish in the $\lambda\rightarrow 0$ 
limit while the poles with a negative imaginary part for their residues
combine in this limit into the projector onto the graded symmetric 
representation of $SU(2|2)$. This suggests that, similarly to non-simply-laced 
affine Toda theories \cite{CDSa, PDorey2}, the poles with 
indefinite-sign residue -- though simple --
should not be interpreted as corresponding to bound states. I would like 
to thank Patrick Dorey for pointing out this possibility. 
} 
The main difficulty with performing the
analysis above for the finite-$\lambda$ S-matrix of \cite{Besu22}
comes from the fact that the symmetry algebra is no longer manifestly
realized.  It is however trivial to see that the S-matrix continues --
for a finite range of $\lambda$ -- to have physical poles at the
expected  position $u_2-u_1=i$. While the residue is no longer a
projector onto the 2-index graded-symmetric representation of
$SU(2|2)$, this symmetry is inherited from the full S-matrix and this
continues to be realized (though not manifestly).

\section{Scattering of higher charge world sheet solitons}

The solitons corresponding to large charge bound states have at least
two representations.
They can be explicitly written as classical solutions of the world sheet 
sigma model in which one relaxes the level-matching condition \cite{HoMa,
MTT3, SpVo, KRT}. Classically, the sigma model can be mapped into the
complex sine-Gordon theory \cite{Dorey2}. The solitons of this theory carry
nonvanishing charge with respect to the $U(1)$ symmetry of this model
and have been argued to correspond to multi-magnon bound
states. There is a one-to-one relation between the complex sine-Gordon 
$U(1)$ charge and the number of bound magnons.

We will first recall the results of \cite{dev2, DoHo} and then
translate them to the variables corresponding to the world sheet sigma
model and compare the result with the large 't~Hooft coupling limit of
the scattering matrices derived in the previous section.

\subsection{Classical scattering of solitons}

To find the classical scattering phase of solitons it is necessary to
construct the solution describing their scattering, extract the time
delay accumulated in the scattering process and then integrate
it. The construction of arbitrary-charge scattering solutions directly
in the world sheet sigma model is quite complicated \cite{SpVo}. 

Classically, the $S^3$ sigma model can be mapped to the Complex
sine-Gordon theory. In particular, there is a one to one
correspondence between the scattering solutions of CsG solitons and
those of the sigma model.  This enables us to use existing results on
their classical scattering phase; we only need to carefully translate
to the dispersion relation of the sigma model. 

Similarly to the scattering solutions of the sigma model, the analytic
expressions for the solutions of CsG are extremely cumbersome, but are
however known for arbitrary charges (see e.g. \cite{dev2} for
details). The time delay was extracted in \cite{dev2, DoHo} from the
large-time asymptotics of these expressions. The CsG solitons are
characterized by their rapidities $\beta$ and by an additional
quantity $\alpha$ related to their $U(1)$ charge. Let us consider two
solitons with rapidities $\beta_{1}$, $\beta_{2}$ and $U(1)$
parameters $\alpha_{1}$ and $\alpha_{2}$. Then, the time delay due to
the collision is \cite{DoHo}:
\begin{equation}
\Delta t(\beta_{1},\beta_{2},\alpha_{1},\alpha_{2})
=\frac{2}{\sinh(\beta_{1})\cos(\alpha_{1})}\log\left |\frac
{\sinh\!\left[\frac{1}{2}(\beta_{1}-\beta_{2})
+\frac{i}{2}(\alpha_{1}-\alpha_{2})\right]}
{\cosh\!\left[\frac{1}{2}(\beta_{1}-\beta_{2})+
\frac{i}{2}(\alpha_{1}+\alpha_{2})\right]}\right |~~.
\label{delay}
\end{equation}
The semi-classical phase shifts $\delta(E_1, E_2)$ for soliton-soliton
scattering is then determined by the WKB formula \cite{JaWo}
\begin{eqnarray}
\frac{d\delta(E_1,\,E_2)}{dE_1}=\Delta t(E_1,\,E_2,\,J_2^{(1)},\,J_2^{(2)})
~~~\longrightarrow~~~
\delta(E_1,\,E_2)=\int dE_1\, \Delta t(E_1,\,E_2,\,J_2^{(1)},\,J_2^{(2)})
~~.
\label{phase_vs_time}
\end{eqnarray}
We need to transform this in variables appropriate to the sigma model, i.e. 
we need to use the bound state dispersion relation rather than the
relativistic dispersion relation of the CsG theory. First,
(\ref{delay}) can be recast into 
\begin{equation}
\Delta t(\beta_{1},\beta_{2},\alpha_{1},\alpha_{2})
=\frac{2}{\sinh(\beta_{1})\cos(\alpha_{1})}\ln \frac
{\cosh(\beta_1-\beta_2)-\cos(\alpha_1-\alpha_2)}
{\cosh(\beta_1-\beta_2)+\cos(\alpha_1+\alpha_2)}~~.
\label{delay1}
\end{equation}
The goal is then to express $\alpha_i$ and $\beta_i$ in terms of the charges $J_2^{(i)}$ and 
the momenta $p_i$. 

As discussed in \cite{Dorey2}, the precise soliton dispersion relation
as well as the expressions of the sigma model momenta and angular
momenta in terms of the CsG parameters are 
\begin{eqnarray}
J_2^{(i)}=2\sqrt{\bar\lambda} \tan\alpha_i\sin^2 \textstyle{\frac{1}{2}}p_i
~,~~~~~~
E_i&=&\sqrt{J_2^{(i)}{}^2+4{\bar\lambda}\sin^2\textstyle{\frac{1}{2}}p_i}
~,~~~~~~
\sinh\beta_i=\frac{\cos\alpha_i}{\tan\textstyle{\frac{1}{2}}p_i}~~.
\label{disp_rel}
\end{eqnarray}
Since the  gauge theory quantities depend only on the square of the 
charges, there is a potential sign ambiguity in the first equation above. For multi-soliton 
configurations it applies for each soliton independently.\footnote{The
charges and energies of each soliton are identified from the
configurations in which the solitons are widely separated.}  
Alternatively, we may choose to fix this ambiguity as in
(\ref{disp_rel}) and interpret solitons with negative charge as
anti-solitons. 

Using (\ref{disp_rel}) implies, after a small amount of algebra, that 
the relevant combinations of  rapidities and $\alpha$-parameters are
\begin{eqnarray}
\cos(\alpha_1\mp\alpha_2)&=&\frac{4{\bar\lambda}
\sin^2{\textstyle{\frac{1}{2}}}p_1\sin^2{\textstyle{\frac{1}{2}}}p_2 \pm J_2^{(1)}J_2^{(2)} 
 }{\sqrt{J_2^{(1)}{}^2+4{\bar\lambda}\sin^4{\textstyle{\frac{1}{2}}}p_1}
   \sqrt{J_2^{(2)}{}^2+4{\bar\lambda}\sin^4{\textstyle{\frac{1}{2}}}p_2}}\cr
\cosh(\beta_1-\beta_2)&=&\frac{\sqrt{J_2^{(1)}{}^2+4{\bar\lambda}
\sin^2{\textstyle{\frac{1}{2}}}p_1}\sqrt{J_2^{(2)}{}^2+4{\bar\lambda}
\sin^2{\textstyle{\frac{1}{2}}}p_2}
-{\bar\lambda} \sin p_1\sin p_2
}
{\sqrt{J_2^{(1)}{}^2+4{\bar\lambda}\sin^4{\textstyle{\frac{1}{2}}}p_1}
\sqrt{J_2^{(2)}{}^2+4{\bar\lambda}\sin^4{\textstyle{\frac{1}{2}}}p_2}}~~.
\end{eqnarray}
Thus, in terms of the sigma model variables, the time delay becomes
\begin{eqnarray}
\label{time_delay}
&&\Delta t=\frac{J_2^{(1)}{}^2+4{\bar\lambda}\sin^4{\textstyle{\frac{1}{2}}}p_1}
{4{\bar\lambda}\sin^3{\textstyle{\frac{1}{2}}}p_1\cos{\textstyle{\frac{1}{2}}}p_1}\times\\
&&	~~~~~~~~~~~~~~
\times\log \frac
{E_1(p_1, J_2^{(1)})E_2(p_2, J_2^{(2)})
-{\bar\lambda} \sin p_1\sin p_2-
(4{\bar\lambda}
\sin^2{\textstyle{\frac{1}{2}}}p_1\sin^2{\textstyle{\frac{1}{2}}}p_2 +
 J_2^{(1)}J_2^{(2)})
}
{
E_1(p_1, J_2^{(1)})E_2(p_2, J_2^{(2)})
-{\bar\lambda} \sin p_1\sin p_2+
(4{\bar\lambda}
\sin^2{\textstyle{\frac{1}{2}}}p_1\sin^2{\textstyle{\frac{1}{2}}}p_2 -
 J_2^{(1)}J_2^{(2)})
}~~.
\nonumber
\end{eqnarray}
Last, using trivially the dispersion relation (\ref{disp_rel}), 
the integration measure in (\ref{phase_vs_time}) becomes
\begin{eqnarray}
\label{measure}
	dE_1=dp_1\,\frac{dE_1}{dp_1}=dp_1\,
	\frac{2{\bar\lambda}\sin{\textstyle{\frac{1}{2}}}p_1
	\cos{\textstyle{\frac{1}{2}}}p_1}
	{\sqrt{J_2^{(1)}{}^2+4{\bar\lambda}\sin^2{\textstyle{\frac{1}{2}}}p_1}}
\end{eqnarray}
With these ingredients we may proceed to compare the classical
soliton-soliton scattering with the predictions of the fused AFS
S-matrix. It is worth mentioning that the scattering phase constructed
from (\ref{time_delay}) and (\ref{measure}) is invariant under the
simultaneous transformation $(J_2^{(1)},\,J_2^{(2)})\rightarrow
(-J_2^{(1)},\,-J_2^{(1)})$, but changes nontrivially under
$(J_2^{(1)},\,J_2^{(2)})\rightarrow (-J_2^{(1)},\,J_2^{(1)})$. 

\section{Comparison \label{comparison}}

Let us now consider separately the two regimes we introduced before:
fixed $J_2^{(i)}$ and  fixed $J_2^{(i)}/\sqrt{\bar\lambda}$ as
$\lambda\rightarrow\infty$. Following the philosophy described in the
introduction, we may obtained the former from the latter as the
limit of vanishing $J_2^{(i)}/\sqrt{\bar\lambda}$. In this sense the
fixed charge magnon bound state may be interpreted as a classical
solution of the sigma model. 

For finite $J_2^{(i)}/\sqrt{\bar\lambda}$ we will discuss
separately the dressing phase  
and complete AFS S-matrix; as mentioned in \S\ref{various_S_matrices},
in both cases we will find a sigma model interpretation. 

\subsection{Large ${\bar\lambda}$ fixed $J_2^{(i)}$}

It is quite trivial to see that in this limit (with fixed $p_i$) all $J_2^{(i)}$-dependence 
becomes irrelevant. Indeed, the expression for the time delay $\Delta t$ can easily be written in terms of 
$J_2^{(i)}/\sqrt{\bar\lambda}$, which vanishes in this limit. Thus, the time delay reduces to that
of \cite{HoMa}, implying that the semi-classical phase shift describing the scattering of fixed charge 
magnon bound states is
\begin{eqnarray}
	\delta_{\rm ws}
	&=&\sqrt{\bar\lambda}\int dp_1\,\sin{\textstyle{\frac{1}{2}}}p_1\,
\log\left[\frac
{1-\cos\textstyle{\frac{1}{2}}p_1\cos\textstyle{\frac{1}{2}}p_2
-\sin\textstyle{\frac{1}{2}}p_1\sin\textstyle{\frac{1}{2}}p_2}
{1-\cos\textstyle{\frac{1}{2}}p_1\cos\textstyle{\frac{1}{2}}p_2
+\sin\textstyle{\frac{1}{2}}p_1\sin\textstyle{\frac{1}{2}}p_2}
\right]\cr
&=&2{\sqrt{\bar\lambda}}\left[\left(\cos
{\textstyle{\frac{1}{2}}}p_2-\cos {\textstyle{\frac{1}{2}}}p_1\right) 
\ln\frac{1-\cos\textstyle{\frac{1}{2}}(p_1-p_2)}
{1-\cos\textstyle{\frac{1}{2}}(p_1+p_2)}-p_1\sin{\textstyle{\frac{1}{2}}}p_2\right]~~.
\label{HM_simple}
\end{eqnarray}

As in \cite{HoMa}, the apparent difference between the prediction of
the fusion construction and  
that of CsG theory may be ascribed to a change of gauge from the
perhaps more standard uniform gauge 
-- if we were to write Bethe equations for states constructed out of
magnon bound states.\footnote{It is worth emphasizing that 
Bethe equations constructed out of the bound state scattering matrix
describe  fewer states than the Bethe equations for elementary magnons.} 
The difference cancels (trivially) between the left- and the
right-hand sides of the Bethe equations.  
This is because the difference between the string length as inherited from the CsG 
analysis and that in the uniform gauge is the energy of the solution
which can further be written as the  
sum over the various (giant) magnons.

\subsection{The general case}

The analysis of the general limit 
\begin{eqnarray}
{\bar\lambda}\rightarrow\infty~~~~~{\rm with}~~~~
{\cal J}_2^{(i)}=J_2^{(i)}/\sqrt{\bar\lambda}={\rm fixed}
\end{eqnarray}
is somewhat more complicated. The goal is to test whether the soliton 
scattering phase constructed in the previous section
\begin{eqnarray}
\label{generalCsG}
&&
\delta^{J_2^{(1)},J_2^{(2)}}_{\rm ws}=\frac{1}{2}
\int \frac{dp_1}{\sin^2{\textstyle{\frac{1}{2}}}p_1}\,
\frac{J_2^{(1)}{}^2+4{\bar\lambda}\sin^4{\textstyle{\frac{1}{2}}}p_1}
{\sqrt{J_2^{(1)}{}^2+4{\bar\lambda}\sin^2{\textstyle{\frac{1}{2}}}p_1}}\times
\\
&&
\times\log \frac
{
\sqrt{J_2^{(1)}{}^2+4{\bar\lambda}\sin^2{\textstyle{\frac{1}{2}}}p_1}
\sqrt{J_2^{(2)}{}^2+4{\bar\lambda}\sin^2{\textstyle{\frac{1}{2}}}p_2}-
4{\bar\lambda}
	\sin{\textstyle{\frac{1}{2}}}p_1\sin{\textstyle{\frac{1}{2}}}p_2
	\cos{\textstyle{\frac{1}{2}}}(p_1-p_2)  -
	 J_2^{(1)}J_2^{(2)}
}
{
\sqrt{J_2^{(1)}{}^2+4{\bar\lambda}\sin^2{\textstyle{\frac{1}{2}}}p_1}
\sqrt{J_2^{(2)}{}^2+4{\bar\lambda}\sin^2{\textstyle{\frac{1}{2}}}p_2}
-4{\bar\lambda}
	\sin{\textstyle{\frac{1}{2}}}p_1\sin{\textstyle{\frac{1}{2}}}p_2
	\cos{\textstyle{\frac{1}{2}}}(p_1+p_2) -
	 J_2^{(1)}J_2^{(2)}
}
\nonumber
\end{eqnarray}
is related to equation (\ref{AFS_full}). 

It turns out that it is possible to directly compute the integral above. To this end 
we first notice that the argument of the logarithm in
(\ref{generalCsG}) may be written as
\begin{eqnarray}
\frac
{
E_1E_2- J_2^{(1)}J_2^{(2)}-
4{\bar\lambda}
	\sin{\textstyle{\frac{1}{2}}}p_1\sin{\textstyle{\frac{1}{2}}}p_2
	\cos{\textstyle{\frac{1}{2}}}(p_1-p_2)  
}
{
E_1E_2-J_2^{(1)}J_2^{(2)}
-4{\bar\lambda}
	\sin{\textstyle{\frac{1}{2}}}p_1\sin{\textstyle{\frac{1}{2}}}p_2
	\cos{\textstyle{\frac{1}{2}}}(p_1+p_2) 
}=
	\frac{1-\frac{{\bar\lambda}/4}{ x^{+(-J_2^{(1)})}x^{-(J_2^{(2)})}}}
	     {1-\frac{{\bar\lambda}/4}{ x^{+(-J_2^{(1)})}x^{+(J_2^{(2)})}}}
	\frac{1-\frac{{\bar\lambda}/4}{ x^{-(-J_2^{(1)})}x^{+(J_2^{(2)})}}}
	     {1-\frac{{\bar\lambda}/4}{
x^{-(-J_2^{(1)})}x^{-(J_2^{(2)})}}}~.
\label{ident0}
\end{eqnarray}
This identity  may be established either analytically or
numerically.  The definition of $x^{\pm(-J)}$ is given by the second equal sign in the
second equation (\ref{full_charges}) with the replacement
$J\rightarrow -J$:
\begin{eqnarray}
x^{\pm(-J)}(p)
=\frac{e^{\pm\frac{i}{2}p}}{ 4\sin\frac{p}{2}}
\left(-J+\sqrt{J^2+4{\bar\lambda}\sin^2\frac{p}{2}}\right)~~.
\end{eqnarray}

Furthermore, we also notice that the coefficient of the
logarithm under the integral sign in equation (\ref{generalCsG}) is a
simple total derivative: 
\begin{eqnarray}
2\frac{d}{dp_1}(u_{J_2^{(2)}}-u_{J_2^{(1)}})=\frac{1}{2\sin^2{\textstyle{\frac{1}{2}}}p_1}\,
\frac{J_2^{(1)}{}^2+4{\bar\lambda}\sin^4{\textstyle{\frac{1}{2}}}p_1}
{\sqrt{J_2^{(1)}{}^2+4{\bar\lambda}\sin^2{\textstyle{\frac{1}{2}}}p_1}}
\end{eqnarray}
Thus, integrating by parts we immediately obtain the first line in
(\ref{AFS_full}) with the transformation  
$J_2^{(1)}\rightarrow -J_2^{(1)}$.

The remaining integral can also be evaluated, though with somewhat more effort. The final 
answer (which may be trivially checked by differentiating with respect to $p_1$ and then 
comparing numerically with the integrand in (\ref{generalCsG})) turns out to be 
\begin{eqnarray}
\delta_{\rm ws}^{-J_2^{(1)},\,J_2^{(2)}}=\delta_{\rm
ws}^{J_2^{(1)},\,-J_2^{(2)}}= \theta_{0}^{J_2^{(1)} J_2^{(2)}}
-p_1\left(\sqrt{J_2^{(2)}{}^2+4{\bar\lambda}
\sin^2{\textstyle{\frac{1}{2}}}p_2}-J_2^{(2)}\right)~~.
\label{dressing_phase_match}
\end{eqnarray}

This result may be interpreted in two different ways. 

1) If we choose to make use of the sign ambiguity in the relation
between the $J_2$ and the CsG parameter $\alpha$ (see below equation
(\ref{disp_rel})), then we may freely change the sign of one of the
charges on the left hand side of the equation above. Thus, with this
prescription, the sigma model scattering of charge-$J_2^{(1)}$ and
charge-$J_2^{(2)}$ solitons appears to reproduce the semiclassical dressing
phase. 

2) If we choose to relate both $J_2^{(1)}$ and $J_2^{(2)}$ and the corresponding CsG
parameters $\alpha_i$ as in the equation (\ref{disp_rel}), then the
sigma model soliton-{\it anti{\hphantom{{$\,$}}}}soliton scattering reproduces the
semiclassical dressing phase. 

In both instances, the difference between the left and
right-hand-sides is the natural extension to finite  
charge states of the analogous term in (\ref{HM_simple}) and as in
that case goes away at the level of the Bethe  
equations.\footnote{This emphasizes that, since S-matrices are
gauge-dependent, two of them cannot be directly  
compared unless they are computed in the same gauge. 
Rather, one should compare physical quantities. Some of the gauge
dependence disappears at the  level of the Bethe equations, which is
what is used in \cite{HoMa} and here.} 

\

The BDS-like part of the gauge theory bound state scattering
matrix may be included in this comparison by making use of further
identities. Indeed, it is not hard to show that 
\begin{eqnarray} 
&&
\frac{1-\frac{{\bar\lambda}/4}{ x^{+(-J_2^{(1)})}x^{-(J_2^{(2)})}}}
	     {1-\frac{{\bar\lambda}/4}{ x^{+(-J_2^{(1)})}x^{+(J_2^{(2)})}}}
	\frac{1-\frac{{\bar\lambda}/4}{ x^{-(-J_2^{(1)})}x^{+(J_2^{(2)})}}}
	     {1-\frac{{\bar\lambda}/4}{
x^{-(-J_2^{(1)})}x^{-(J_2^{(2)})}}}
=\\
&&~~~~~~~~~~~~~~~~
=
\frac{1-\frac{{\bar\lambda}/4}{ x^{+(J_2^{(1)})}x^{-(J_2^{(2)})}}}
        {1-\frac{{\bar\lambda}/4}{ x^{+(J_2^{(1)})}x^{+(J_2^{(2)})}}}
\frac{1-\frac{{\bar\lambda}/4}{ x^{-(J_2^{(1)})}x^{+(J_2^{(2)})}}}
        {1-\frac{{\bar\lambda}/4}{x^{-(J_2^{(1)})}x^{-(J_2^{(2)})}}}
\times
\frac{(u_{J_2^{(1)}}-u_{J_2^{(2)}})^2+{\textstyle{\frac{1}{4}}}(J_2^{(1)}-J_2^{(2)})^2}
       {(u_{J_2^{(1)}}-u_{J_2^{(2)}})^2+{\textstyle{\frac{1}{4}}}(J_2^{(1)}+J_2^{(2)})^2}
~~.
\nonumber
\end{eqnarray}
This implies that, with the relation between both $J_2^{(1)}$ and
$J_2^{(2)}$ and the corresponding CsG parameters $\alpha_i$ as in
equation (\ref{disp_rel}), the sum of the first lines of equations
(\ref{AFS_full}) and (\ref{potential_extra}) reproduces the result of
the integration by parts in equation (\ref{generalCsG}). 

As before, it is possible (with some effort) to perform the
remaining integral with the result 
\begin{eqnarray}
\delta_{\rm ws}^{J_2^{(1)},\,J_2^{(2)}}=\theta_{0}^{J_2^{(1)}
J_2^{(2)}}
+\delta\theta_{0}^{J_2^{(1)} J_2^{(2)}}
-p_1\left(\sqrt{J_2^{(2)}{}^2+4{\bar\lambda}
\sin^2{\textstyle{\frac{1}{2}}}p_2}-J_2^{(2)}\right)~~.
\label{w_quantum}
\end{eqnarray}
Therefore, up to the same term accounting for the difference of gauge
choice, the sigma model scattering  phase reproduces both the dressing
phase as well as contribution of the fused BDS S-matrix. 

%
%
%

\section{The small momentum limit and further checks}

The two results of the previous section appear somewhat surprising: 
both the complete AFS scattering matrix and the dressing phase have
independent world sheet interpretation. They are the soliton-soliton
and soliton-antisoliton scattering phases. We will now provide independent
evidence that this is indeed correct. 

To this end we will consider the limit in which the world sheet
soliton become regular string states:
\begin{eqnarray}
J_2^{(i)}\rightarrow 1~~~~~~~~p\rightarrow 0~~~~~~{\bar\lambda
p^2}={\rm fixed}~~. 
\end{eqnarray}
In this limit we should be able to compare (up to gauge and coordinate
artifacts) the scattering phases (\ref{dressing_phase_match}) and
(\ref{w_quantum}) with the scattering amplitudes of the world sheet
fields corresponding to gauge theory magnons. 

It is relatively clear that $\delta_{\rm ws}^{J_2^{(1)},
J_2^{(1)}}$  becomes, in this limit, the scattering of the holomorphic
world sheet scalars of the Landau-Lifshitz model \cite{KZ, RTT2}.  A 
potential subtlety relates  to the
fate of $\delta_{\rm  ws}^{-J_2^{(1)}, J_2^{(1)}}$ in that
it should be related to the scattering fields of opposite
R-charge. Due to its nonrelativistic nature, the world sheet image 
of the gauge theory $SU(2)$ sector -- the Landau-Lifshitz model -- 
capture only the scattering of likewise charge fields.  Thus, the
scattering amplitude which should match $\delta_{\rm ws}^{-1, 1}$
lies outside the $SU(2)$ sector.\footnote{This conclusion should in fact not be
unexpected. Up to gauge and coordinate artifacts the dressing phase
is universal and thus its world sheet interpretation needs not be
restricted to a single sector.}  Consequently, to identify on the world sheet 
the low momentum and low angular momentum limit of dyonic giant 
magnon scattering we should analyze the sigma model without
restricting the fields to their positive energy modes. Relaxing this
restriction allows fields of opposite R-charges in the initial state. 
In the full string theory the scattering of neutral configurations (such 
as solitons and antisolitons) leads in the final state to (essentially) 
all possible neutral combination of fields. Thus, testing e.g. the 
factorization of the S-matrix necessarily requires (some of) the other 
fields of the theory.
The soliton-antisoliton scattering phase obtained from the CsG theory 
captures {\it only} the ``exclusive'' amplitude with the same final 
and initial states.

The dyonic giant magnons have, by construction, an infinitely large
angular momentum $J_1$. It appears therefore natural to use, similarly
to \cite{AFuniform},  a gauge fixing it. The complete semiclassical world 
sheet scattering matrix in this gauge will be discussed elsewhere \cite{KMRZ}. 
The action restricted to $R\times S^3$ is easy to
construct following the strategy of \cite{KAT}. 
Starting with the $R\times S^3$ metric 
\begin{eqnarray}
ds^2=-dt^2 + \frac{(1-y{\bar y})^2}{(1+y{\bar y})^2}\,d\phi^2 
+\frac{4dyd{\bar y}}{(1+y{\bar y})^2}~~,
\end{eqnarray}
performing  2d duality along $\phi$ and fixing the uniform gauge
$t=\tau$ and ${\tilde\phi}=\frac{J_1}{\sqrt{\lambda}}\sigma$,
redefining the spatial coordinate $\sigma = \frac{2\pi}{J_1}x$ and then scaling  $J_1$ to
infinity we obtain an action defined on the plane 
($S=\int d\tau \int_0^\infty dx L$) and the Lagrangian
\begin{eqnarray}
L &=& \frac{1}{2}
\left[
(\partial_\tau y)^2-{\bar\lambda}(\partial_x y)^2-y^2
\right]\cr
&+&\frac{1}{2}\left({\bar\lambda}({\partial_x y})^2-({\partial_\tau
y})^2\right)\left({\bar\lambda}({\partial_x {\bar y}})^2 -
({\partial_\tau {\bar y}})^2\right) +{2}{\bar\lambda}\,y {\bar y}\,
{\partial_x y}\,{\partial_x {\bar y}} 
-\frac{1}{2} y^2 {\bar y}^2+\dots~~,
\label{lag_unif}
\end{eqnarray}
where the ellipsis stands for terms with more than six fields. 
With the mode expansion
\begin{eqnarray}
y=\int\frac{dk_1}{\sqrt{2 k_0}}~\big(\,a(k)e^{-ik\cdot z}+b(k){}^\dagger
e^{+ik\cdot z}\,\big)~~~,~~~~~k_0\equiv e(k_1)=\sqrt{1+{\bar\lambda}k_1^2}
\end{eqnarray}
which leads to canonical commutation relations,  the amplitude of the process
$yy\rightarrow {y}{y}$ is \cite{KMRZ}
\begin{eqnarray}
S_{y(p_1)y(p_2)\rightarrow {y}(p_1){y}(p_2)}
&=&\frac{1}{4}\frac{1}{p_1e(p_2)-p_2e(p_1)}
\left[2\left({\bar\lambda}p_1p_2 -
e(p_1)e(p_2)\right)^2+2{\bar\lambda}(p_1+p_2)^2-2\right]\cr
&=&-(p_1-p_2)+(\theta_{0}^{11}+\delta\theta_{0}^{11})
\big|_{p_i\rightarrow 0 ~~~
\atop {\bar\lambda} p_i^2={\rm fixed}}~~\vphantom{{}^{{}^{\big|}}}
\end{eqnarray}
is  the same (up to the change of gauge) as in \cite{RTT2}. Thus, in
the small momentum limit, the soliton-soliton scattering phase becomes
the scattering phase of the world sheet fields corresponding to the
``elementary'' magnons. 

The amplitude of the process $y(p_1){\bar y}(p_2)\rightarrow
{y}(p_1){\bar y}(p_2)$ can be equally well extracted from
(\ref{lag_unif}). It is \cite{KMRZ}
\begin{eqnarray}
S_{y(p_1){\bar y}(p_2)\rightarrow {y}(p_1){\bar y}(p_2)}
&=&\frac{1}{4}\frac{1}{p_1e(p_2)-p_2e(p_1)}
\left[2\left({\bar\lambda}p_1p_2 -
e(p_1)e(p_2)\right)^2+2{\bar\lambda}(p_1-p_2)^2-2\right]\cr
&=&-(p_1-p_2)+\theta_{0}^{11}
\Big|_{p_i\rightarrow 0 ~~~
\atop {\bar\lambda} p_i^2={\rm fixed}}~~\vphantom{{}^{{}^{\big|}}}
\end{eqnarray}
which, up to the gauge and coordinate artifacts leading to the first
term above and the last term in (\ref{dressing_phase_match}), agrees
with our expectations and confirms the identification of the sigma
model $(-J_2^{(1)},J_2^{(2)})$ soliton-antisoliton scattering phase 
with the fused dressing phase.  

\section{Summary and Discussions}

In an integrable quantum field theory the scattering of the bound
states is determined by the S-matrix 
of its constituents. If independent calculations are available for
some of its limits, the comparison  tests the consistency of the
scattering matrix of the constituents of the bound state. 

We have re-analyzed the boostrap construction without assuming
Lorentz-invariance and found the conditions  
that a simple pole of the scattering matrix describes a physical bound
state. Formally  it is unchanged from the equivalent condition in a
Lorentz-invariant theory. Using this condition we have  
constructed the scattering matrix of bound states of arbitrary (fixed or/and scaling) 
$J_2$ charge under the assumption that their constituents scatter with the BDS 
and AFS-type S-matrices. The analytic structure of the result is
somewhat richer than that of  the ``elementary'' magnons,
exhibiting both simple and double-poles.  

We compared the result based on the AFS-type magnon S-matrix with the
scattering of the corresponding  world sheet solitons. Their
connection to the complex sine-Gordon theory provides easy access to
their  scattering phase. 

\noindent
$\bullet$ We found that the semiclassical scattering phase
reproduces, with the identification of charges of the scattering
solitons as in equation (\ref{disp_rel}), the complete bound state
scattering matrix,  incorporating both the fused BDS factor and
the dressing phase.  This corresponds to the first standpoint described in
\S\ref{various_S_matrices}. 
Through the 't~Hooft coupling dependence of the fused BDS factor (which 
appears to be of a one-loop origin), this result provides an additional 
test of the integrability of the quantum world sheet theory. This is in 
the same spirit as the comparison of the one-loop world sheet corrections 
to the energies of extended semiclassical strings and finite size 
corrections from the gauge theory Bethe equations \cite{BeTZ}.
 
The equation (\ref{w_quantum}) 
suggests that a possible approach to the calculation of the quantum
corrections to the elementary magnon  
scattering matrix is to first find the quantum corrections to the
large charge soliton-soliton  
scattering and discretize them. There are at least three seemingly 
different ways to approach such a calculation. On the one hand, one
may use a moduli space approximation.  
To this end it is necessary to construct the sigma model scattering
solution (perhaps along the lines  
of \cite{SpVo}), find the quantum corrections to the moduli space
metric and then read off the corrections  
to the scattering amplitude. This approximation is clearly restricted
to the small momentum limit. 
On the other hand, one may find the 1PI effective 
action of the world sheet sigma model and then simply search for
solutions of its equations of  
motion which reduce in the classical limit to the initial scattering
solutions and then use the  
WKB approximation. Alternatively, one may try to reformulate the world
sheet sigma model such that the new fields  
correspond to solitons.\footnote{In the case of the (real) sine-Gordon
theory this reformulation is  
the Thirring model.} Then, in this theory the quantum corrections we
are interested in would be responsible  
for the double-poles in the bound state scattering matrix. It would be
interesting to see if these approaches lead to consistent answers
which are in agreement with those of \cite{HeLo}.

\noindent
$\bullet$ We have also found that the dressing phase by itself has a
sigma model interpretation, the precise details of which depend on
one's choice of identification of soliton charges with CsG parameters. 
On the one hand, by making use of the ambiguity in the identification 
of the sigma model angular momenta $J_2$ and CsG parameters and
choosing a different identification for the two solitons, it is
possible to adjust them such that the dressing phase (without the fused
BDS contribution) is reproduced by the sigma model scattering. 
On the other hand,  if we universally fix the identification of the sigma model  
angular momenta $J_2$ and CsG parameters as in \cite{Dorey2}, then the
dressing phase is reproduced  by the sigma model soliton-{\it
anti\hphantom{{$\,$}}}soliton  scattering. These two
observations are in fact related  by the inability to
differentiate  between an isolated soliton
and an isolated anti-soliton.  

These observations, independently confirmed in the small momentum
limit, follow the second standpoint described 
in \S\ref{various_S_matrices}   that, based on their 't~Hooft coupling
dependence, the fused BDS S-matrices should be considered of  
a quantum origin even though their $J_2$ dependence makes them scale 
semiclassically. It would be interesting to explicitly check
whether quantum corrections to the classical  soliton-antisoliton
scattering indeed reproduce the contribution of the fused BDS phase. 
Since the soliton-soliton and soliton-antisoliton scattering matrices 
are related -- in the complex sine-Gordon theory -- by the relativistic 
crossing transformations, the result of such a comparison would give further 
information on the realization of crossing symmetry in the AdS/CFT correspondence.

Though it is included in our discussion, we have not separately analyzed the 
comparison of the scattering phase of a unit charge magnon off a fixed ${\cal J}_2$
bound state. The fact that one of the states has a semiclassical interpretation
may make more tractable the calculation of quantum corrections  to this scattering 
process. More generally, quantum corrections to world sheet solitons appear to 
be of importance for their relation to the gauge theory side of the AdS/CFT 
correspondence. It is possible that quantum corrections to the Hofman-Maldecena soliton
will teach us about their scattering off ``elementary'' magnons.

\

\

\

\noindent
{\bf Acknowledgments:}

I would like to thank Benasque Center for Physics, the Niels
Bohr Institute and SPhT Saclay for hospitality 
while this work was in progress and the organizers of the 
Albert Einstein Institut workshop on Integrability in Gauge and 
String Theory for a stimulating atmosphere.  
It is a pleasure to thank Nick Dorey,  Sergey Frolov, Tristan McLoughlin, 
Joe Minahan, Kostya Zarembo for useful discussions.
I would especially like to thank Arkady Tseytlin for illuminating 
discussions and comments on the draft.
%

\end{document}